\documentstyle[12pt]{article}
\begin{document}
\title{ {\normalsize
         \begin{tabbing}
            \'DESY 92-055 \`ISSN 0418-9833\\
            \'April 1992 \\
           \\
         \end{tabbing}}
Non-Critical Strings in\\
       Robertson-Walker Space Time}

\author{K. Behrndt\thanks{bitnet: behrndt@znsun1.ifh.de}\\
{\normalsize \em DESY-Institut f\"ur Hochenergiephysik, Zeuthen}}
\date{\ }
\maketitle
\begin{abstract}
I consider a D+1 dimensional nonlinear $\sigma$ model based on a possible
interpretation of the Liouville field as a physical time. The Weyl invariance
of this theory gives us restrictions for the background fields and
the parameters of the theory, e.g.\ for
trivial background one obtains the known regions for the dimension of the
space-time ($\leq$1 or $\geq$25). For a Robertson-Walker space time a
special solution of these equations is discussed.
\end{abstract}

\renewcommand{\arraystretch}{1.6}
\renewcommand{\thefootnote}{\alph{footnote}}

\section{Introduction}
The nonlinear $\sigma$ model describing strings in non non-trivial
background fields was extensively studied in the last years \cite{1,2}.
By the interpretation of the
Liouville field as physical time it is possible to regard the
non-critical string
theory as a special $\sigma$ model \cite{3,4,5}. Some interesting results for
non-trivial target space were already obtained. First of all this is
the black hole solution  for $D = 1$ \cite{3,12,16}. Secondly it is possible
to quantize the theory in arbitrary dimension if one incorporates convenient
spatial background charges \cite{19}.
And finally the new cosmological results are remarkable
\cite{4,10,13}.

In this paper I consider a string living in a Robertson-Walker space time
where the time correspond to the Liouville field
and I want discuss a special solution of the Weyl invariance conditions.
The following questions found my  particular interest: 1) How is the
timelike evolution of the metric?  2) What is the region of the allowed
dimension of the space time? and 3) Is it possible to find a solution
for the tachyon equation? Similar questions are already discussed for
flat  (spatial independent) Robertson-Walker metrics \cite{5}.

\section{The Model}
Before I want consider the string in a Robertson-Walker space time
let me make some introductory remarks for the theory in Minkowskean
(or Euclidean) space time. For genus zero world sheets the corresponding
partition function is given by \cite{7,8} :
\begin{equation}
Z=\int D_{\hat{g}}\sigma\,D_{\hat{g}}{\bf X}\ e^{-S}
\end{equation}
with:\begin{equation}
      \begin{array}{l}
  S( {\bf X} , \sigma , \hat{g} ) = S_{M}({\bf X} , \hat{g}) + \frac{1}{4\pi}
    \int d^2 z \sqrt{\hat{g}} \left( \hat{g}^{ab}
    \partial_a \sigma \partial_b \sigma + Q \hat{R} \sigma + \mu
    e^{2\alpha\sigma} \right) \\
   S_{M}( {\bf X},\hat{g} ) = \frac{1}{4\pi\alpha'} \int d^2 z \sqrt{\hat{g}}\,
     \hat{g}^{ab} \partial_a {\bf X} \cdot \partial_b {\bf X} \ .
  \end{array}
\end{equation}
Here ${\bf X}$ is the position of the string in the space time and $\sigma$
is the Liouville field connected with the 2d metric via the conformal
gauge: $g_{ab} = e^{2\sigma} \hat{g}_{ab}$ ; $\hat{g}_{ab}$ is a reference
metric and $\hat{R}$ denotes the 2d scalar curvature with respect to the
reference metric $\hat{g}_{ab}$. The coefficients $Q$ and $\alpha$ in $S$
are not arbitrary \cite{8}: $Q$ is determinated by the vanishing of the total
central charge and $\alpha$ by the requirement that the conformal dimension
of $e^{2\alpha\phi}$ is equal to one.

If one interprets the Liouville field as a time variable it is possible
to consider this model as a special $\sigma$ model \cite{3,4,5}.
With \[
\begin{array}{ccc}
D \rightarrow D + 1 \ \ &,&\ \  \sigma \rightarrow \frac{1}{\sqrt{\alpha'}} X^0
\end{array}
\]
one obtains
\begin{equation}
  \begin{array}{l}
    Z = \int D_{\hat{g}}X \  e^{-S} \ \ , \\
    S = \frac{1}{4\pi\alpha'} \int_M d^2 z \sqrt{\hat{g}}\left(\hat{g}^{ab}
     \partial_a X^{\mu} \partial_b X^{\nu} G_{\mu\nu} + \alpha' \hat{R} \phi
     + \alpha' T \right)\ .
  \end{array}
\end{equation}
Here $G_{\mu\nu}$ is the metric in the target space (space time), $\phi$
is the dilaton field and $T$ the tachyon field. For the special background
fields: $G_{\mu\nu} = \delta_{\mu\nu}$, $\phi(X^0) =\frac{1}{\sqrt{\alpha'}}
QX^0$ and $T(X^0) = \mu e^{2\frac{\alpha}{\sqrt{\alpha'}}X^0}$
one obtains again the original model (1), (2).

Since the functional integration includes the integration over the Liouville
field this $\sigma$ model must be Weyl invariant. This statement is equivalent
to the vanishing of the total central charge \cite{8}. As the original theory
depends on $g_{ab}$ only (not on $\hat{g}_{ab}$), this statement seems to be
trivial. But if we regularize this theory we have to introduce a cutoff
corresponding to $\hat{g}_{ab}$ and therefore the condition for
Weyl invariance give us non-trivial restrictions for the background fields
\cite{4}. The vanishing of the Weyl anomaly correspond to the vanishing
of all $\bar{\beta}$ functions of the theory \cite{1}:
\begin{equation}
 \begin{array}{llr}
 & \bar{\beta}^i \equiv 0 \ \ \ \ \forall i\ \  &\mbox{( i numerates all
                                    background fields )} , \\
\mbox{with}:\ \ \ \ & \bar{\beta}^{G}_{\mu\nu} =
                  \beta^{G}_{\mu\nu} + D_{(\mu}M_{\nu)}
             \ \ ,\ \ & M_{\nu} = 2\alpha' \partial_{\nu} \phi + W_{\nu}\ ,\\
    &  \bar{\beta}^{\phi} = \beta^{\phi} +
                          \frac{1}{2} M^{\mu}\partial_{\mu}\phi\ , &\\
    &  \bar{\beta}^T = \beta^T - 2T + \frac{1}{2} M^{\mu}\partial_{\mu}T \ . &
 \end{array}
\end{equation}
Up to the second order in $\alpha'$ one gets for the $\beta$
functions \cite{1,11}\footnote{I neglect in my
consideration all `` non-perturbative '' contributions \cite{4,11}.}:
\begin{equation}
 \begin{array}{l}
  \beta^T = -\frac{1}{2} \alpha' D^2 T  \ , \\
  \beta^{G}_{\mu\nu} = \alpha' R_{\mu\nu} + \frac{1}{2} \alpha'^2
     R_{\mu\alpha\beta\lambda}R_{\nu}^{\ \alpha\beta\lambda}  \ ,\\
   \beta^{\phi} = \frac{1}{6}(D-25) - \frac{1}{2}\alpha' D^2 \phi +
       \frac{1}{16} \alpha'^{2}(R_{\mu\nu\alpha\lambda})^2 \ .
 \end{array}
\end{equation}
\vspace{5mm}

In the following I want to investigate these equations and look for
non-trivial solutions. As a first example I want rederive the
results for $Q$ and $\alpha$ obtained by David, Distler and Kawai \cite{8}.
Therefore I consider the background field configuration:
\begin{equation}
  G_{\mu\nu} = \delta_{\mu\nu} \ \ \ ,\ \ \  \phi = \frac{1}{\sqrt{\alpha'}}
   Q X^0 \ \ \ \ \mbox{and}\ \ \
  T = \mu e^{2\frac{\alpha}{\sqrt{\alpha'}} X^0} \ .
\end{equation}
As a result of the covariance $W_{\nu}$ in (4) is a function of
the covariant derivatives of the curvature tensor only (torsion terms
are absent in this model) and it would vanish if these curvature terms
vanished, especially $W_{\nu}$ is independent from the dilaton and tachyon
field \cite{1}. For the equations (4) one obtains:
\begin{equation}
 \begin{array}{l}
   \bar{\beta}^{G}_{\mu\nu} \equiv 0 \ ,\\
   \bar{\beta}^{\phi} = \frac{1}{6}(D-25) + Q^2 = 0 \ ,\\
   \bar{\beta}^T = -2( \alpha^2 - Q\alpha + 1 )T = 0\ .
 \end{array}
\end{equation}
Thus we have the known results:
\begin{equation}
 \begin{array}{rcl}
    Q &=& \sqrt{\frac{25-D}{6}}\\
    \alpha_{\pm} &=& \frac{Q}{2} \pm \sqrt{\frac{Q^2}{4} - 1} =
      \frac{1}{\sqrt{24}} \left(\sqrt{25-D} \pm \sqrt{1-D} \right) \ .
 \end{array}
\end{equation}
\vspace{5mm}

Before I turn to a non-trivial
space time metric I look for the most general dilaton and tachyon
field in {\em flat} space time ($G_{\mu\nu} = \delta_{\mu\nu}$) solving
the equations (4). In this case one has:
\begin{equation}
 \bar{\beta}^{G}_{\mu\nu} = 2\alpha' \partial_{\mu}\partial_{\nu} \phi = 0
\end{equation}
and hence
\begin{equation}
 \phi = \phi_0 + \frac{1}{\sqrt{\alpha'}} q_{\mu} X^{\mu}\ ,
\end{equation}
which was already discussed as one example for noncritical strings
in arbitrary dimensions \cite{19}. The tachyon field is given by:
\begin{equation}
 \bar{\beta}^T = - \frac{1}{2}\alpha'\partial^2 T - 2T + \sqrt{\alpha'}
      q_{\mu}\partial^{\mu}T = 0 \ ,
\end{equation}
with the solution:
\begin{equation}
 T \sim e^{\frac{1}{\sqrt{\alpha'}}p_{\mu}X^{\mu}} \hspace{10mm}
  \mbox{and:} \hspace{10mm}
  -\frac{1}{2}p_{\mu}p^{\mu} - 2 + q_{\mu}p^{\mu} = 0\ .
\end{equation}
For the dilaton $\bar{\beta}$ function one gets:
\begin{equation}
 \bar{\beta}^{\phi} = \frac{1}{6}(D-25) + q_{\mu}q^{\mu} = 0 \ .
\end{equation}
Combining (12) and (13) one obtains finally \cite{6,20}:
\begin{equation}
 \begin{array}{l}
   q_0 = \sqrt{\frac{25-D}{6} - \vec{q}^{\;2}} \ , \\
 p_{0_{\pm}} = q_0 \pm  \sqrt{q_0^{2} - \vec{p}^{\;2} + 2\vec{q}\vec{p} - 4} =
   \frac{1}{\sqrt{6}} \left( \sqrt{25-D - 6\vec{q}^{\;2} } \pm
   \sqrt{ 1-D - 6(\vec{q} - \vec{p})^2 } \right) \ .
 \end{array}
\end{equation}

\section{A Solution for Robertson-Walker Space Time}
Now I want to investigate the equations (4) in a Robertson-Walker
space time. This metric describing a (spatial) homogeneous and isotropic
universe is given by \cite{17}:
\begin{equation}
  (ds)^2 = a^2 (dX^{0})^{2} + \frac{K^2 (X^0)}{(1 + \frac{1}{4}\epsilon r^2)^2}
       \left[ (dX^1)^2 + (dX^2)^2 + ... + (dX^D)^2 \right] \ ,
\end{equation}
where $r^2 = (X^1)^2 + (X^2)^2 + ... + (X^D)^2$ , $K(X^0)$ is the so-called
world radius and the parameter $\epsilon$ determinates the spatial geometry:
flat ($\epsilon = 0$), spherical ($\epsilon = +1$) or hyperbolical
($\epsilon = -1$)\footnote{Strictly speaking $\epsilon$ has to carry
space time dimensions and therefore $\epsilon = \pm \frac{1}{\alpha'} , 0$.}.
This metric has Euclidean signature if $a^2 > 0$ and Minkowskean
signature if $a^2 < 0$. The only non-vanishing curvature terms are given
by ($k,l,... = 1,...,D$):
\begin{equation}
  R^{m}_{\ nkl} = \left[ \frac{\epsilon}{K^2} - \frac{1}{a^2}
    \left(\frac{\dot{K}}{K}\right)^2 \right] \left( G^{m}_{\ k} G_{nl} -
    G^{m}_{\ l} G_{nk} \right) \ \ \ ,\ \ \
  R^{0}_{\ n0l} = -\frac{1}{a^2} \frac{\ddot{K}}{K} G_{nl}\ .
\end{equation}
If one computes the $\beta$ functions in terms of this curvature tensor
one obtains very complicated non-linear differential equations for
$K$, $\phi$ and $T$. In order to avoid these difficulties I will not consider
$G_{\mu\nu}$ and $\phi$ as ``physical'' metric and dilaton \cite{1,10},
but (for $D \neq 1$)\footnote{By this
reparametrization $S_{eff} = \int d^{D+1}X
\sqrt{G} e^{-2\phi} \left[ \frac{2}{3}(D-25) - \alpha' R - 4\alpha'
(\partial \phi)^2 + ... \right]$ is for $D \neq 1$
just transformed in $S_{eff} =
\int d^{D+1} X \sqrt{\tilde{G}} \left[ \alpha' \tilde{R} + 2\alpha'
(\partial \tilde{\phi})^2 + \frac{2}{3}(D-25)e^{\frac{2}{\sqrt{|D-1|}}
\tilde{\phi}} + ... \right]$,
i.e.\ an action with a standard kinetic part for the dilaton and the usual
volume measure. In addition the theory in $\tilde{G}_{\mu\nu}$ and
$\tilde{\phi}$ yield the correct vertex operators for the $S$-matrix
\cite{1}.}:
\begin{equation}
  \tilde{G}_{\mu\nu} = e^{-\frac{4}{D-1}\phi} G_{\mu\nu} \ \ \ ,\ \ \
  \tilde{\phi} = \frac{2}{\sqrt{|D-1|}} \phi \ .
\end{equation}
Therefore it is reasonable that the time evolution of the metric
is controlled by the dilaton field, i.e.:
\begin{equation}
  K = \mbox{const.} \ \ ,\ \  \phi = \phi ( X^0 )\ \ .
\end{equation}
Then one gets for (16):
\begin{equation}
 R_{mnkl} = \frac{\epsilon}{K^2} \left( G_{mk} G_{nl} -
      G_{ml} G_{nk} \right) \ \ \ ,\ \ \ R_{0n0l} = 0 \ .
\end{equation}
This configuration corresponds for $\epsilon = 0$ or $D = 1$
to a flat space time and in addition all covariant derivatives
and timelike components of the curvature tensor vanish. Hence
it is impossible to construct a vector field $W_{\mu}$
or a timelike tensor $\beta_{00}$ (they have to be independent of the
dilaton and tachyon field \cite{1}) and thus:
$W_{\mu} = \beta^{G}_{00} = 0$, $M_{\mu} \equiv 2 \alpha'
\partial_{\mu} \phi$.
So one finds for the metric and dilaton $\bar{\beta}$ functions (4) in
arbitrary orders in $\alpha'$:
\begin{equation}
 \begin{array}{lr}
  \bar{\beta}^{G}_{00} = 2 \alpha' D_0 \partial_0 \phi(X^0) = 2 \alpha'
                       \ddot{\phi} \ , & \\
  \bar{\beta}^{G}_{mn} = \beta^{G}_{mn}  & (D_m \partial_n \phi = 0 \
                \mbox{if}\  K = \mbox{const.\ and}\  \partial_m \phi = 0) \ ,\\
  \bar{\beta}^{\phi} = \beta^{\phi} + \frac{\alpha'}{a^2}
  \left( \dot{\phi}(X^0) \right)^2 \ . &
 \end{array}
\end{equation}
Note that from (19) follows: $\beta^{G}_{mn} = \Lambda G_{mn}$ and the
vanishing of $\bar{\beta}^{G}_{mn}$ is equivalent to the vanishing of
$\Lambda$ which restrict the possible value of $K$ and $\epsilon$ (presumed
it exist real solution) and furthermore the vanishing of the metric
$\beta$ function ensure that one does not need renormalize the metric.
The vanishing of the first equation gives us:
\begin{equation}
 \begin{array}{lr}
  \phi(X^0) = \frac{a}{\sqrt{\alpha'}} Q X^0 + \phi_0\ \ \ ;\ \  &
 \mbox{and\ with\ the\ last\ equation}:\ \ \ Q = \pm \sqrt{ - \beta^{\phi} }\ .
 \end{array}
\end{equation}
Here $Q$ is a generalized background charge which depend from the $\epsilon$
and the dimension of the space time and $\phi_0$ is the zero mode
($\sqrt{\alpha'}$ ensure that $Q$ is dimensionless and $a$ that
$Q$ is a scalar under time reparametrization). As for the dilaton field
I want investigate at this point a tachyon field depending of the time only.
Than $\beta^T$ can not contain curvature terms ( $R_{0n0l} = 0$ ) and one finds
for $\beta^T$ the same equation as in the flat case. In terms of
(4) $T$ is defined by the equation:
\begin{equation}
  - \frac{1}{2} \frac{\alpha'}{a^2} \ddot{T}(X^0) - 2 T(X^0) +
    \frac{\alpha'}{a^2} \dot{\phi} \dot{T}(X^0) = 0 \ ,
\end{equation}
with the solution:
\begin{equation}
  T (X^0) \sim e^{2 \frac{\alpha_{\pm}}{\sqrt{\alpha'}} a X^0} \hspace{10mm}
  \mbox{wherein:} \ \
  \alpha_{\pm} = \frac{1}{2} \left( Q \pm \sqrt{ Q^2 - 4} \right)
\end{equation}
or with (21):
\begin{equation}
  \alpha_{\pm} = \frac{1}{2} \left( \sqrt{ -\beta^{\phi}} \pm
  \sqrt{-\beta^{\phi} - 4} \right)\ .
\end{equation}
Via $\beta^{\phi}$  in this equation one has again (cp.\ (5)) a restriction
for the dimension of the space time. The tachyon field is real if:
$\beta^{\phi} \leq -4$ (Euclidean space time) or
$\beta^{\phi} \geq 0$ (Minkowskean space time). For $D = 1$ or $\epsilon = 0$
all curvature terms vanish and thus for $\epsilon = 0$:
$\beta^{\phi} = \frac{1}{6} (D-25)$; or for $D = 1$: $\beta^{\phi} = -4$.
In addition the metric $\beta$ function vanish and one does not
have any restriction for $K$.
\vspace{5mm}

All expressions I have obtained up to this point are valid in arbitrary
order in $\alpha'$. Now I compute $Q$ up to the second order
and want discuss the modification for a spatial dependent tachyon field.
In terms of (5) one finds:
\begin{equation}
 \begin{array}{l}
  \beta^{G}_{mn} = \left[ (D-1) \frac{\alpha' \epsilon}{K^2} + (D-1) \left(
          \frac{\alpha' \epsilon}{K^2} \right)^2 \right] G_{mn} \ ,   \\
  \beta^{\phi} = \frac{1}{6} (D-25) + \frac{D(D-1)}{8} \left(
           \frac{\alpha' \epsilon}{K^2} \right)^2 \ .
 \end{array}
\end{equation}
Therefore one gets for the vanishing of the metric $\beta$
function the restrictions:
\begin{equation}
 \begin{array}{l}
  \hspace{20mm}  K^2 = - \alpha' \epsilon \hspace{10mm} \mbox{or} \hspace{10mm}
     \epsilon = 0 \hspace{10mm}  \mbox{or} \hspace{10mm} D = 1 \\
    \mbox{and\ thus\ if\ } D \neq 1\ \mbox{and}\ \epsilon \neq 0 : \\
     \hspace{20mm} Q = \sqrt{-\beta^{\phi}} = \sqrt{ \frac{25-D}{6} -
    \frac{D(D-1)}{8}} \ .
 \end{array}
\end{equation}
In order to obtain a {\em real} tachyon field (23), (24) we get the
restriction for Euclidean signature ($a^2 > 0$): $Q^2 \geq 4$ or
$-\frac{4}{3} \leq D \leq 1$ and
for Minkowskean signature ($a^2 < 0$): $Q^2 \leq 0$ which correspond to
 $D \geq \frac{1}{6}
(\sqrt{1201}-1) \approx 5.61$ or $D \leq -\frac{1}{6}(\sqrt{1201}+1)$.
Unfortunately the dependence on $\alpha'$ in $Q$ drops out and
hence one has not a perturbative result. Therefore one can not neglect
the next terms and one has to deal with this result carefully.

A spatial dependent tachyon field is up to the second order
defined by:
\begin{equation}
 -\frac{1}{2} \alpha' D^2 T(X^0,X^m) - 2 T(X^0,X^m) + \frac{\alpha'}{a^2}
 \dot{\phi} \dot{T}(X^0,X^m) = 0\ .
\end{equation}
With the ansatz: $T(X^0 , X^m) = H(X^m) U(X^0)$ and the assumption that $H$
should depend on $L = \frac{1}{1+\frac{1}{4} \epsilon r^2}$ only (motivated
by the spatial dependence of $G_{\mu\nu}$), one can for $\epsilon \neq 0$
separate this equation in two ones ($\lambda$ is the separation constant):
\begin{equation}
 \begin{array}{cr}
  - \frac{1}{2} \frac{\alpha'}{a^2} \ddot{U} + \frac{\sqrt{\alpha'}}{a} Q
    \dot{U} - \lambda U = 0 \ , &\\
  L(1-L) H''(L) + ( \frac{D}{2} - D L ) H'(L) + 2(2-\lambda) H(L) = 0 \ &
   (\ H' = \frac{d}{d L} H\ )\ .
 \end{array}
\end{equation}
The solution for $U$ is given by (23) if one replaces $\alpha_{\pm}$
by $\alpha_{\pm} = \frac{1}{2}(Q \pm \sqrt{Q^2 - 2\lambda})$. The
second equation is just the hypergeometrical differential equation
and solutions around $L = 1$ $(r^2 \simeq 0)$, around $L = 0$
$(r^2 \simeq \infty)$ and for $\alpha' \epsilon = -1$
around $L \simeq \infty$ ($\alpha' r^2 \simeq 4$) can one finds in \cite{15}.
Hypergeometrical functions as correlation functions for conformal field
theories in restricted domains are already discussed in \cite{18}.
For $\epsilon = 0$
one gets for the tachyon by replacing from $p_0 \rightarrow 2ap_0$ and
$p_m \rightarrow 2Kp_m$ and setting $q_{\mu} = 0$ just the flat solution
(12), (14).
\vspace{5mm}

Finally I return to the results (21)-(24) and transform they in
the `` physical'' one. In order to interpret $\tilde{G}_{\mu\nu}$ in (17)
as ``physical'' metric from Robertson-Walker type a time reparametrization
$X^0 \rightarrow t = t(X^0)$
transforming $\tilde{G}_{00}$ in $\pm 1$ ($+1$ for Euclidean and $-1$
for Minkowskean signature) is quite reasonable. Of course this procedure
makes sence only for $D \neq 1$. The corresponding
transformation is given by:
\begin{equation}
 \begin{array}{c}
  a X^{0}_{\pm} = \frac{\sqrt{\alpha'} (D-1)}{2 Q} \log \frac{ (D-1)
  \sqrt{\alpha'}} {2 \tilde{Q} \sqrt{\eta} \, ( t_0 \pm t )} \\
 \ \hfill (\tilde{Q} = Q e^{\frac{2 \phi_0}{D-1}}\ , \ t_0 = \mbox{const.}\ ,\
    \eta = \pm 1\ \mbox{for\ Minkowskean\ or\ Euclidean\ signature}\ ) \ .
 \end{array}
\end{equation}
If one prefers for $t$ the same direction as for $X^0$ one has to take
the minus sign.  After performing this transformation one finds finally
for the ``physical'' metric, dilaton field and tachyon field:
\begin{equation}
 \begin{array}{l}
  (d s)^2 = \eta (d t)^2 + \frac{\tilde{K}^2(t)}{(1 + \frac{1}{4}
  \epsilon r^2)^2} \left[ (d X^1)^2 + (d X^2)^2 + ... + (d X^D)^2 \right] \ ,\\
   \tilde{\phi} = \frac{\sqrt{|D-1|}}{2} \log \frac{K^2}{\tilde{K}^2}
   \hspace{10mm} \mbox{and} \hspace{10mm} \tilde{T}(t)
    \sim \left( \frac{K^2}{ \tilde{K}^2} \right)^{\delta_{\pm}}\ ,\\
   \mbox{with}:\hspace{10mm}
    \tilde{K}^2 = \frac{4 \eta Q^2 }{\alpha' (D-1)^2 } (t \pm t_0)^2 K^2
    \hspace{10mm}
    \mbox{and} \hspace{10mm} \delta_{\pm} = \frac{\alpha_{\pm} (D-1)}{2Q} \ .
 \end{array}
\end{equation}
Here for $\epsilon = 0$ $K^2$ is arbitrary and $Q = \sqrt{\frac{D-25}{6}}$,
for $\epsilon = \pm \alpha'$
$K^2$ is given by the equation: $\beta_{mn} = 0$, $Q$ is given by (21) and
$\alpha_{\pm}$ by (24). Up to the order $\alpha'^2$ the corresponding
expressions  are given in equation (26).

\section{Conclusion}
I have discussed solutions of the Weyl invariance condition for a
string $\sigma$ model in which the time correspond to the Liouville
field. The vanishing of the Weyl anomaly coefficients ($\bar{\beta}$
function) restricts the possible background fields and the parameters of
the theory. First I have rederived the known results for $Q$ and
$\alpha_{\pm}$ (David, Distler,Kawai). Secondly I investigated
the general solution in a {\em flat} space time. The corresponding
dilaton field must be linear and the tachyon field is just a exponential
function. The zero components of the momentums are restricted, but the
dimension is for special regions of spatial momentums (or
spatial background charge resp.) arbitrary.

As an example for non-trivial metric I have studied
in the third section a solution for a Robertson-Walker space time.
If one transforms the metric and dilaton in the ``physical'' ones
($\tilde{G}_{\mu\nu}$, $\tilde{\phi}$) and furthermore assumes
that the time evolution is controlled by the dilaton field
it is possible to find solutions for the Wyel invariance condition
(also in higher orders in $\alpha'$). The value for the background
charge and for $\alpha_{\pm}$ is in principle now given by the
dilaton $\beta$ function. The demand that the tachyon should
be a real field restricts the dilaton $\beta$ function and thus
the dimension: for Euclidean space time $\beta^{\phi} \leq -4$ and
for Minkowskean $\beta^{\phi} \geq 0$. For $\epsilon = 0$ or $D =1$:
$\beta^{\phi} = -4$ and else one has perturbative results only.
The obtained results are valid for arbitrary order in $\alpha'$.
A perturbative computation of $\beta^{\phi}$ or the generalized
background charge resp.\ $Q$ yields up to the second order in
$\alpha'$ the restriction: $-\frac{4}{3} \leq D \leq 1$ (Euclidean
signature) or $D \geq \frac{1}{6}(\sqrt{1201} - 1) \approx 5.61$
or $D \leq -\frac{1}{6}(\sqrt{1201} + 1)$. Unfortunately the
parameter of the perturbation theory $\alpha'$ drops out and one
has to deal with this result carefully.

The  final result for the ``physical'' background fields solving the Weyl
invariance condition is given by (30). The world radius
depends quadratic, the dilaton logarithmic and the tachyon field
potential on the time. The direction of the evolution (increasing or
decreasing) is determined by $\epsilon$ and the signature of the space time
(Euclidean or Minkowskean).
\vspace{6mm}

\noindent
{\large \bf Acknowledgments} \vspace{3mm}

I would like to thank H.\ Dorn and J.\ Schnittger for very useful discussions.

\renewcommand{\arraystretch}{1}

\end{document}